\documentclass[showpacs,prd,twocolumn,amsmath,amssymb]{revtex4}
\usepackage{graphicx}% Include figure files
\usepackage{dcolumn}% Align table columns on decimal point
\usepackage{bm}% bold math
\usepackage{slashed}
\usepackage{feyn}
\def\bra{\langle}
\def\ket{\rangle}
\def\d{\partial}
\newcommand{\R}{\mathbb{R}}
\newcommand{\cA}{\mathcal{A}}
\newcommand{\cD}{\mathcal{D}}
\newcommand{\cH}{\mathcal{H}}
\newcommand{\cN}{\mathcal{N}}
\def\cG{\mathcal{G}}
\def\cP{\mathcal{P}}

\def\Ei{\mathrm{Ei}}
\def\bvec#1{\mathbf{#1}}
\def\dk#1#2{\frac{ d^{#2}{#1} }{ (2\pi)^{#2} }} % invariant measure in FT

\def\eg{{\sl e.g. }}

\def\ie{{\sl i.e. }}

\def\vb{\bvec{b}}
\def\vk{\bvec{k}}
\def\vq{\bvec{q}}
\def\vp{\bvec{p}}
\def\vx{\bvec{x}}
\def\vy{\bvec{y}}

\begin{document}

\title{Continuous Wavelet Transform in Quantum Field Theory}% Force line breaks with \\

\author{M.V.Altaisky}
\affiliation{Space Research Institute RAS, Profsoyuznaya 84/32, Moscow, 117997, Russia}
\email{altaisky@mx.iki.rssi.ru}
\altaffiliation[Also at ]{Bogoliubov Laboratory of Theoretical Physics, Joint Institute for Nuclear Research, Dubna, 141980, Russia}%Lines break automatically or can be forced with \\
\author{N.E.Kaputkina}
\affiliation{National Technological University ''MISiS'', Leninsky prospect 4, Moscow, 119049, Russia}
\email{nataly@misis.ru}
%\homepage{http://lrb.jinr.ru/People/altaisky/MVAltaiskyE.html}

\date{revised June 15, 2013}% It is always \today, today,
             %  but any date may be explicitly specified

\begin{abstract}
We describe the application of the continuous wavelet transform to calculation of the Green functions in quantum field theory: scalar $\phi^4$ theory, quantum electrodynamics, quantum chromodynamics. The method of continuous wavelet transform in quantum field theory presented in \cite{Altaisky2010PRD} for the scalar 
$\phi^4$ theory, consists in substitution of the local fields $\phi(x)$ by those dependent on both the position $x$ and the resolution $a$. 
The substitution of the action $S[\phi(x)]$ by the action $S[\phi_a(x)]$ makes the local theory into nonlocal one, and implies the causality conditions related to the scale $a$, the {\em region causality} 
\cite{CC2005}. These conditions make the Green functions 
$G(x_1,a_1, \ldots, x_n,a_n)= \bra \phi_{a_1}(x_1)\ldots \phi_{a_n}(x_n)\ket
$
 finite for any given set of regions by means of an effective cutoff scale $A=\min (a_1,\ldots,a_n)$.     
\end{abstract}

% PACS, the Physics and Astronomy
                             % Classification Scheme.
\maketitle

\section{Introduction}
The fundamental problem of quantum field theory and statistical 
mechanics is the problem of divergences of  Feynman integrals 
emerging in Green functions. 
The formal infinities appearing in perturbation 
expansion of Feynman integrals are tackled with different regularization methods,
from Pauli-Villars regularization to renormalization methods for 
gauge theories, see \eg \cite{Collins1984} for a review. 
A special class of regularizations are the lattice regularizations tailored for the precise numerical simulations 
in gauge theories \cite{Kogut1979,BMMP2010}.

There are a few basic ideas connected with those
regularizations. First, certain minimal scale $L=\frac{2\pi}{\Lambda}$, where $\Lambda$ is the cutoff momentum, is introduced into the theory, with all
the fields $\phi(x)$ being substituted by their Fourier
transforms truncated at momentum $\Lambda$.
%$$
%\phi(x)\to\phi_{\left(\frac{2\pi}{\Lambda}\right)}(x) = \int_{|k|\le\Lambda} e^{-\imath k x} \tilde \phi(k) \dk{k}{d}.
%$$
%(The same happens on the lattice but through the discrete transform.)
The physical quantities are then demanded to be independent on the
rescaling of the cut-off parameter~$\Lambda$. The second thing is the Kadanoff blocking
procedure~\cite{Kadanoff1966}, which averages the small-scale fluctuations up to a certain scale --
this makes a kind of effective interaction.

Physically all these methods imply to the self-similarity assumption:
blocks interact to each other similarly to the sub-blocks \cite{GK1983}. Similarly, but
not necessarily having the same interaction strength -- the latter can be
dependent on scale $\lambda =\lambda(a)$.
However there is no place for such dependence
if the fields are described solely in terms of their Fourier transform -- except for the cutoff momentum dependence. The latter representation, being based on the representation of 
the translation group, is rather restrictive: it determines the effective interaction of {\em all fluctuations up to a certain scale}, but says nothing
about the interaction of the fluctuations at a given scale.

%%%%%%%%%%%%%%%%%%%%%%%%%%%%%%%%%%%%%%
That is why, the functional methods capable of taking into 
account the interaction at specific scale are required. 
Wavelet analysis, being the multiscale alternative to 
the Fourier transform, emerged in geophysics \cite{MAFG1982}, is the 
most known of such methods. Its application to quantum field theory have been suggested by many authors \cite{BS1994,HS1995,Federbush1995,Battle1999,Alt2002G24J}
The other side of the problem is that the quantum nature of the 
fields considered in quantum field theory is constrained 
by the Heisenberg uncertainty principle. 
To localize a particle in an interval $\Delta x$ the measuring device 
requests a momentum transfer of order $\Delta p\!\sim\!\hbar/\Delta x$. If $\Delta x$ is too small the field $\phi(x)$ at 
a fixed point $x$ has no experimentally verifiable meaning. What is meaningful, is the vacuum expectation of product of fields in certain 
region centered around $x$, the width of which ($\Delta x$) is constrained 
by the experimental conditions of the measurement \cite{Altaisky2010PRD}. That is why, at least from physical 
point of view, any such field should be designated by the 
resolution of observation $\phi_{\Delta x}(x)$.
  
In present paper we exploit the observation that quantum field theory models, 
which yield divergent 
Feynman graphs, can be studied analytically if we project original fields 
$\phi(x)$ into the fields $\phi_a(x)$, subscribed by the scale of measurement 
$a$. The Green functions $\bra \phi_{a_1}(x_1)\ldots\phi_{a_n}(x_n)\ket$ become 
finite under certain causality assumptions, which stand for the fact that 
any $n$-point correlation function can be dependent only on space-time regions, 
rather than points, and thus cannot be infinite \cite{Altaisky2010PRD}. 
These Green function describe the effect of propagation of a perturbation from  
a region of size $a$, centered at a point $x$, to a region of size $a'$, centered at a point $x'$. The standard quantum field theory models can be reformulated 
by expressing the (point-dependent) local fields $\phi(x)$, the distributions, in terms of the 
region-dependent fields $\phi_a(x)$. The integration over all scales $a$ will 
of course drive us back to the known divergent results, but the physical observables are always those measured with finite resolution, and their correlations 
are always finite. Therefore the idea of wavelet transform of quantum fields, 
which will be considered below, is very similar to the idea of renormalization 
group (This similarity, being studied in the lattice framework
\cite{Battle1999,BP2013}, is out of coverage of the present paper). 

The remainder of this paper is organized as follows. In 
{\em Section ~\ref{cwt:sec}} we recall the basics of the continuous wavelet transform and its application to the 
multiresolution analysis of quantum fields. The definitions 
of scale-dependent fields and Green functions, the modifications of the Feynman diagram technique are 
presented. The $\phi^4$ scalar field model examples of calculations are given. {\em Section ~\ref{op:sec}} 
considers the case of operator-valued scale-dependent 
fields. The operator ordering and commutation relations 
are presented. The relations between the theory of scale-dependent fields in Euclidean and Minkowski spaces 
are discussed. {\em Section ~\ref{gau:sec}} presents the 
examples of calculation of one-loop Feynman graphs in QED 
and QCD. The {\em Conclusion} gives a few remarks on the 
perspectives and applicability of the multiscale field theory 
approach based on continuous wavelet transform.   
  
%%%%%%%%%%%%%%%%%%%%%%%%%%%%%%%%%%%%%%%%%%%%%%%%%%%%%%%%%%%%%%
\section{Continuous wavelet transform in quantum field theory \label{cwt:sec}}
\subsection{Basics of the continuous wavelet transform}
Let $\cH$ be a Hilbert space of states for a quantum field $|\phi\ket$. 
Let $G$ be a locally compact Lie group acting transitively on $\cH$, 
with $d\mu(\nu),\nu\in G$ being a left-invariant measure on $G$. Then, 
similarly to representation of a vector $|\phi\ket$ in a Hilbert space 
of states $\cH$ as a linear combination of an eigenvectors of momentum 
operator 
$
|\phi\ket=\int |p\ket dp \bra p |\phi\ket,$
any $|\phi\ket \in \cH$ can be decomposed with respect to 
a representation $U(\nu)$ of $G$ in $\cH$ \cite{Carey1976,DM1976}:
\begin{equation}
|\phi\ket= \frac{1}{C_g}\int_G U(\nu)|g\ket d\mu(\nu)\bra g|U^*(\nu)|\phi\ket, \label{gwl} 
\end{equation} 
where $|g\ket \in \cH$ is referred to as an admissible vector, 
or {\em basic wavelet}, satisfying the admissibility condition 
$$
C_g = \frac{1}{\| g \|^2} \int_G |\bra g| U(\nu)|g \ket |^2 
d\mu(\nu)
<\infty. %\label{adc}
$$
The coefficients $\bra g|U^*(\nu)|\phi\ket$ are referred to as 
wavelet coefficients. 

If the group $G$ is abelian, the wavelet transform \eqref{gwl} with 
$G:x'=x+b'$ coincides with Fourier transform. 

The next to the abelian group is the group of the affine transformations 
of the Euclidean space $\R^d$
\begin{equation}
G: x' = a R(\theta)x + b, x,b \in \R^d, a \in \R_+, \theta \in SO(d), \label{ag1}
\end{equation} 
where $R(\theta)$ is the rotation matrix.
We define unitary representation of the affine transform \eqref{ag1} with 
respect to the basic wavelet $g(x)$ as follows:
\begin{equation}
U(a,b,\theta) g(x) = \frac{1}{a^d} g \left(R^{-1}(\theta)\frac{x-b}{a} \right).
\end{equation}  
(We use $L^1$ norm \cite{Chui1992,HM1996} instead of usual $L^2$ to keep the physical dimension 
of wavelet coefficients equal to the dimension of the original fields).

Thus the wavelet coefficients of the function $\phi(x) \in L^2(\R^d)$ with 
respect to the basic wavelet $g(x)$ in Euclidean space $\R^d$ can be written 
as 
\begin{equation}
\phi_{a,\theta}(b) = \int_{\R^d} \frac{1}{a^d} \overline{g \left(R^{-1}(\theta)\frac{x-b}{a} \right) }\phi(x) d^dx. \label{dwtrd}
\end{equation} 
The wavelet coefficients \eqref{dwtrd} represent the result of the measurement 
of function $\phi(x)$ at the point $b$ at the scale $a$ with an aperture 
function $g$ rotated by the angle(s) $\theta$ \cite{PhysRevLett.64.745}.

The function $\phi(x)$ can be reconstructed from its wavelet coefficients 
\eqref{dwtrd} using the formula \eqref{gwl}:
\begin{equation}
\phi(x) = \frac{1}{C_g} \int \frac{1}{a^d} g\left(R^{-1}(\theta)\frac{x-b}{a}\right) \phi_{a\theta}(b) \frac{dad^db}{a} d\mu(\theta) \label{iwt}
\end{equation}
The normalization 
constant
$C_g$ is readily evaluated using Fourier transform:
%\begin{equation}
$$
C_g = \int_0^\infty |\tilde g(aR^{-1}(\theta)k)|^2\frac{da}{a} d\mu(\theta)
= \int |\tilde g(k)|^2 \frac{d^dk}{|k|^d}<\infty.
$$
%\label{adcf}
%\end{equation}
For isotropic wavelets 
%\begin{equation}
$$
C_g = \int_0^\infty |\tilde g(ak)|^2\frac{da}{a}
= \int |\tilde g(k)|^2 \frac{d^dk}{S_{d}|k|^d},
$$
%\label{adcfi}
%\end{equation}
where $S_d = \frac{2 \pi^{d/2}}{\Gamma(d/2)}$ is the area of unit sphere 
in $\R^d$.
\subsection{Resolution-dependent fields}
If the ordinary quantum field theory defines the field 
function $\phi(x)$ as a scalar product of the state vector of the system 
 and the state vector corresponding to the localization at the point $x$:
\begin{equation}
\phi(x) \equiv \bra x | \phi \ket,
\end{equation}
the modified theory \cite{AltSIGMA07,Altaisky2010PRD} should respect the resolution of the measuring equipment. Namely, we define the 
{\em resolution-dependent fields} 
\begin{equation}
\phi_{a\theta}(x) \equiv \bra x,\theta, a; g|\phi\ket,\label{sdf}
\end{equation}
also referred to as the scale components of $\phi$,
where $\bra x, \theta, a; g|$ is the bra-vector corresponding to localization 
of the measuring device around the point $x$ with the spatial resolution $a$ and the orientation $\theta \in SO(d)$;
$g$ labels the apparatus function of the equipment, an {\em aperture}
\cite{PhysRevLett.64.745}. The field theory of extended objects with the basis $g$  defined on the spin variables was considered in \cite{GS2009,Varlamov:2012}.

The introduction of resolution into the definition of 
the field function has a clear physical interpretation. 
If the particle, described by the field $\phi(x)$, have been 
initially prepared in the interval 
$(x-\frac{\Delta x}{2},x+\frac{\Delta x}{2})$, the probability of 
registering this particle in this interval is generally less than unity:
for the probability of registration depends on the strength of interaction 
and the ratio of typical scales of the measured particle and the measuring 
equipment. The maximum probability of registering an object of 
typical scale $\Delta x$ by the equipment with typical resolution $a$
 is achieved when these two parameters are comparable. For this reason 
the probability of registering an electron by visual range photon scattering 
is much higher than by that of long radio-frequency waves. As 
mathematical generalization, we should say that if a measuring equipment  
with a given spatial resolution $a$ fails to register an object, prepared 
on spatial interval of width $\Delta x$ with certainty, 
then tuning the equipment to {\em all} possible resolutions $a'$ would 
lead to the registration. This certifies the fact of the existence 
of the measured object.

In terms of the resolution-dependent 
field \eqref{sdf} the unit probability of registering the object $\phi$ 
anywhere in space at any resolution and any orientation of the measuring device
 is expressed by normalization  
\begin{equation}
\int |\phi_{a,\theta}(x)|^2d\mu(a,\theta,x)=1,
\end{equation}
where $d\mu(a,\theta,x)$ is an invariant measure on $\R_+ \times SO(d)\times \R^d$, 
which depends on the position $x$, the resolution $a$, and the orientation $\theta$ of the aperture $g$.

If the measuring equipment has the resolution $A$, \ie all 
states $\bra g;a\ge A,x|\phi\ket$ are registered with significant 
probability, but those with $a<A$ are not, the regularization of the 
field theory in momentum space, 
with the cutoff momentum $\Lambda=2\pi/A$ corresponds to the 
UV-regularized functions
\begin{equation}
\phi^{(A)}(x) =\frac{1}{C_g}\int_{a\ge A} 
\bra x|g;a,b\ket d\mu(a,b)\bra g;a,b|\phi\ket.
\end{equation}
The regularized $n$-point Green functions are 
$
\mathcal{G}^{(A)}(x_1,\ldots,x_n) \equiv \bra\phi^{(A)}(x_1),\ldots, \phi^{(A)}(x_n) \ket_c 
.$

However, the momentum cutoff is merely a technical trick: the physical 
analysis, performed by renormalization group method \cite{HooftVeltman1972,Wilson1973,Collins1984}, demands 
the independence of physical results from the cutoff at $\Lambda\to\infty$.

\subsection{Scalar field example}
To illustrate the method, following \cite{AltSIGMA07,Altaisky2010PRD},  we start 
with Euclidean scalar field theory. The widely known example 
which fairly illustrates 
the problem is the $\phi^4$ interaction model in 
$\R^d$, see \eg \cite{Collins1984,Ramond1989}, determined by the generating functional 
\begin{equation}
W[J]= \cN \int  
e^{-\int d^dx \left[ 
\frac{1}{2}(\d\phi)^2+\frac{m^2}{2}\phi^2 + \frac{\lambda}{4!}\phi^4
 - J\phi \right]  } \cD \phi, \label{gf1}
\end{equation}
where $\cN$ is a formal normalization constant.
The connected Green functions are given 
by variational derivatives of the generating functional:
\begin{equation}
G^{(n)} = 
\left. { \frac{\delta^n\ln W[J]}{\delta J(x_1) \ldots \delta J(x_n)}
}\right|_{J=0}.
\label{cgf}
\end{equation}
In statistical sense these functions  have the meaning of 
the $n$-point correlation functions \cite{ZJ1999}.
The divergences of Feynman graphs in the perturbation expansion 
of the Green functions \eqref{cgf} with respect to the  
coupling constant $\lambda$  emerge at coinciding arguments 
$x_i=x_k$. For instance, the bare two-point correlation 
function 
\begin{equation}
G^{(2)}_0(x-y) = \int \frac{d^dp}{(2\pi)^d}\frac{e^{-\imath p(x-y)}}{p^2+m^2}
\end{equation}
is divergent at $x\!=\!y$ for $d\ge2$. 

For simplicity let us assume the basic wavelet $g$ to be isotropic, i.e. 
we can drop the rotation matrix $R(\theta)$. 
Substitution of the continuous wavelet transform \eqref{iwt} into field theory \eqref{gf1}   
gives the generating functional for the scale-dependent fields $\phi_a(x)$ 
\cite{AltSIGMA07}:
\begin{widetext}
\begin{align} \nonumber 
W_W[J_a] &=&\cN \int \cD\phi_a(x) \exp \Bigl[ -\frac{1}{2}\int \phi_{a_1}(x_1) D(a_1,a_2,x_1-x_2) \phi_{a_2}(x_2)
\frac{da_1d^dx_1}{a_1}\frac{da_2d^dx_2}{a_2}  \\
&-&\frac{\lambda}{4!}
\int V_{x_1,\ldots,x_4}^{a_1,\ldots,a_4} \phi_{a_1}(x_1)\cdots\phi_{a_4}(x_4)
\frac{da_1 d^dx_1}{a_1} \frac{da_2 d^dx_2}{a_2} \frac{da_3 d^dx_3}{a_3} \frac{da_4 d^dx_4}{a_4} 
+ \int J_a(x)\phi_a(x)\frac{dad^dx}{a}\Bigr], \label{gfw}
\end{align}
\end{widetext}
with $D(a_1,a_2,x_1-x_2)$ and $V_{x_1,\ldots,x_4}^{a_1,\ldots,a_4}$ denoting the wavelet images of the inverse propagator and that of the interaction potential. 
The Green functions for scale component fields are given by 
functional derivatives
$$
\bra\phi_{a_1}(x_1)\cdots\phi_{a_n}(x_n)\ket_c
= \left. \frac{\delta^n \ln W_W[J_a]}{\delta J_{a_1}(x_1)\ldots 
\delta J_{a_n}(x_n)} \right|_{J=0}.$$
Surely the integration in \eqref{gfw} over all 
scale variables $\int_0^\infty \frac{da_i}{a_i}$ turns us back 
to the divergent theory \eqref{gf1}.

This is the point to restrict the functional integration in \eqref{gfw} 
only to the field configurations $\{ \phi_a(x) \}_{a\ge A}$. The restriction 
is imposed at the level of the Feynman diagram technique. Indeed, applying 
the Fourier transform to the r.h.s. of (\ref{iwt},\ref{dwtrd}) one yields 
\begin{eqnarray*}
\phi(x) &=& \frac{1}{C_g} \int_0^\infty \frac{da}{a} \int \dk{k}{d} e^{-\imath k x}
\tilde g(ak) \tilde \phi_a(k), 
\\  
\tilde\phi_a(k) &=& \overline{\tilde g(ak)}\tilde\phi(k) .
\end{eqnarray*}
Doing so, we have the following modification of the Feynman diagram technique
\cite{Alt2002G24J}:
\begin{itemize}\itemsep=0pt
\item each field $\tilde\phi(k)$ will be substituted by the scale component
$\tilde\phi(k)\to\tilde\phi_a(k) = \overline{\tilde g(ak)}\tilde\phi(k)$.
\item each integration in momentum variable is accompanied by corresponding 
scale integration:
\[
 \dk{k}{d} \to  \dk{k}{d} \frac{da}{a}.
 \]
\item each interaction vertex is substituted by its wavelet transform; 
for the $N$-th power interaction vertex this gives multiplication 
by factor 
$\displaystyle{\prod_{i=1}^N \overline{\tilde g(a_ik_i)}}$.
\end{itemize}
According to these rules the bare Green function in wavelet representation 
takes the form 
%\begin{equation}
$$
G^{(2)}_0(a_1,a_2,p) = \frac{\tilde g(a_1p)\tilde g(-a_2p)}{p^2+m^2}.
$$ 
%\label{g20}
%\end{equation}   
The finiteness of the loop integrals is provided by the following rule:
{\em there should be no scales $a_i$ in internal lines smaller than the minimal scale 
of all external lines}. Therefore the integration in $a_i$ variables is performed from 
the minimal scale of all external lines up to the infinity.

\begin{figure}[ht]
\includegraphics[width=.33\textwidth]{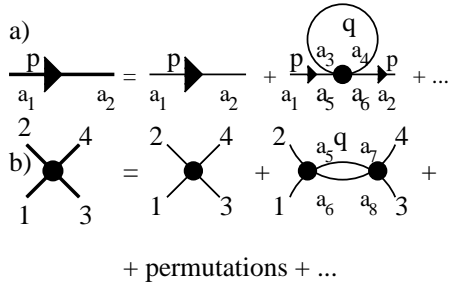}
\caption{Feynman diagrams for the Green functions $G^{(2)}$ and $G^{(4)}$ for the resolution-dependent fields. Redrawn from \cite{Altaisky2010PRD}}
\label{gf:pic}
\end{figure}

To understand how the method works one can look at the one-loop 
contributions to the two-point 
Green function $G^{(2)}(a_1,a_2,p)$ shown in 
Fig.~\ref{gf:pic}a., and to the vertex shown in 
Fig.~\ref{gf:pic}b. 
The best choice of 
the wavelet function $g(x)$ would be the apparatus function of 
the measuring device, however any well localized function with 
$\tilde g(0)=0$ will suit. 
The tadpole integral, to keep with the notation of \cite{AltSIGMA07}, 
is written as  
\begin{align*}\nonumber 
T_1^d(Am)&=&\frac{1}{C_g^2} \int_{a_3,a_4 \ge A}  \frac{d^dq}{(2\pi)^d} 
\frac{|\tilde g(a_3 q)|^2 |\tilde g(-a_4 q)|^2}{q^2+m^2}
 \frac{da_3}{a_3} \frac{da_4}{a_4} \\   
&=&\frac{S_d m^{d-2}}{(2\pi)^d}\int_0^\infty f^2(Amx)\frac{x^{d-1}dx}{x^2+1}, 
\end{align*}
where the integration over the scale variables resulted in the 
effective cutoff function 
\begin{equation}
f(x) \equiv \frac{1}{C_g}\int_x^\infty |\tilde g(a)|^2\frac{da}{a},
\quad f(0)=1, \label{cutf1}
\end{equation} 
%%%%%%%%%%%%%%%%%%%% Ins 1 %%%%%%%%%%%%%
which depends on the squared modulus of the Fourier 
image of the basic wavelet, and thus is even with respect to reflections.
%%%%%%%%%%%%%%%%%%%%%%%%%%%%%%%%%%%%%%%%

In the one-loop contribution to the vertex, shown in Fig.~\ref{gf:pic}b,
the value of the loop integral is 
\begin{equation}
X_d = \frac{\lambda^2}{2}\frac{1}{(2\pi)^d}\int \frac{d^dq}{(2\pi)^d}
\frac{f^2(qA)f^2((q-s)A)}{\left[ q^2+m^2\right]\left[ (q-s)^2+m^2\right] },
\label{I4}
\end{equation}
where $s\!=\!p_1\!+\!p_2, A=\min(a_1,a_2,a_3,a_4)$. The integral \eqref{I4} can 
be calculated by symmetrization of loop momenta $q\!\to\!q\!+\!\frac{s}{2}$ in 
Fig.~\ref{gf:pic}b, introducing dimensionless variable $\vy = \vq/s$,
after a simple algebra  we get   
\begin{align*}\nonumber 
X_d=\frac{\lambda^2}{2}\frac{S_{d-1} s^{d-4} }{(2\pi)^{2d}} 
\int_0^\pi d\theta \sin^{d-2}\theta \int_0^\infty dy y^{d-3}\times  \\
 \frac{ f^2\left(As \sqrt{y^2+y\cos\theta + \frac{1}{4}}\right)
f^2\left(As \sqrt{y^2-y\cos\theta + \frac{1}{4}}\right)
 }{
\left[
\frac{y^2+\frac{1}{4}+\frac{m^2}{s^2}}{y} + \cos\theta
\right]
\left[
\frac{y^2+\frac{1}{4}+\frac{m^2}{s^2}}{y} - \cos\theta
\right]
},
\end{align*}
where $\theta$ is the angle between the loop momentum $q$ and the total 
momentum $s$.
For the simple choice of the basic wavelet $g_1$ \cite{AltSIGMA07,Altaisky2010PRD}
%\begin{equation}
$$
g(x)=-\frac{x e^{-x^2/2}}{(2\pi)^{d/2}},\quad \tilde g(k) = \imath k e^{-k^2/2}
$$
%\label{g1}
%\end{equation}
in four dimensions we get a finite result 
\begin{align*}
T^4_1(\alpha^2) &=& \frac{-4\alpha^4 e^{2\alpha^2} \Ei_1(2\alpha^2)+2\alpha^2
}{
64\pi^2\alpha^4}m^2, \\
\lim_{s^2\gg 4m^2} X_4(\alpha^2) &=& \frac{\lambda^2}{256\pi^6} \frac{e^{-2\alpha^2}}{2\alpha^2} 
\bigl[e^{\alpha^2}-1 
- \alpha^2e^{2\alpha^2}\Ei_1(\alpha^2) \\
&+& 2\alpha^2e^{2\alpha^2}\Ei_1(2\alpha^2)
  \bigr], 
\end{align*}
depending on  is dimensionless scale factor $\alpha\!\equiv\!Am$, 
where $A$ is the minimal scale of all external lines.   

These results display an evident fact that for the massive scalar field 
all length scales are to be measured in the units of inverse mass.

\section{Causality and commutation relations \label{op:sec}}

\subsection{Operator ordering}
Up to now we have considered the calculation of the Feynman diagrams 
for the scale-dependent fields $\phi_{a,\cdot}(x)$ treated as 
$c$-valued functions. In quantum 
field theory, adjusted to high energy physics applications, the fields 
$\phi_{a,\cdot}(x)$ are operator-valued functions. So, as it was 
already emphasized in the context of the wavelet application to 
quantum chromodynamics \cite{Federbush1995,Altaisky2010PRD}, the operator ordering and the commutation relations  are to be defined.

\begin{figure}[ht]
\centering \includegraphics[width=0.3\textwidth]{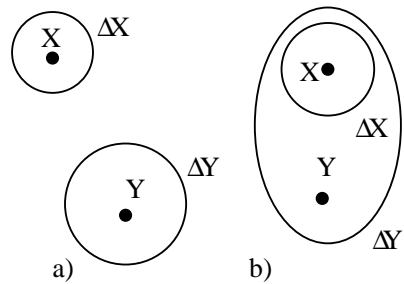}
\caption{Causal ordering of scale-dependent fields. Space-like regions are drawn in Euclidean space: a) The event regions do not intersect; b) Event $X$ is inside the event $Y$.}
\label{tnew2:pic}
\end{figure}

\begin{figure}[ht]
\centering \includegraphics[width=0.3\textwidth]{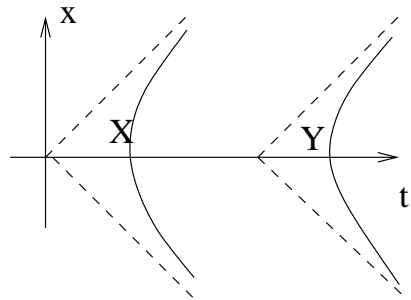}
\caption{Disjoint events in ($t,x$) plane in Minkowski space}
\label{mink1:pic}
\end{figure}

\begin{figure}[ht]
\centering \includegraphics[width=0.3\textwidth]{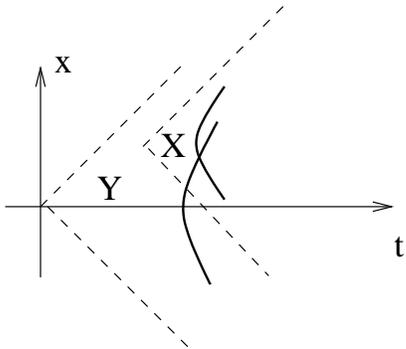}
\caption{Nontrivial intersection of two events $X\subset Y$ in ($t,x$) plane in Minkowski space}
\label{mink2:pic}
\end{figure}

In standard quantum field theory the operator ordering is performed 
according to the non-decreasing of the time argument  in 
the product of the operator-valued functions acting on 
vacuum state
$$
\underbrace{A(t_n)A(t_{n-1}) \ldots A(t_2) A(t_1)}_{t_n \ge t_{n-1}\ge \ldots \ge t_2 \ge t_1}|0\ket. 
$$
In the infinite momentum frame (IMF), which simplifies algebraic structure of the current algebra, the time-ordering is performed in the proper time argument $x_+$ 
\cite{ChangMa1969}. 
The quantization is performed by separating the Fourier transform 
of quantum fields into the positive- and the negative-frequency 
parts
$$
\phi = \phi^+(x) + \phi^-(x),$$ defined as follows
\begin{align}
%\nonumber 
\phi(x) %     &=& \left(\int_0^\infty dk_0  + \int_{-\infty}^0 dk_0\right)\int \frac{d^{d-1} k}{(2\pi)^d}       e^{\imath k x} u(k) \\
        &=& 
\int \frac{d^d k}{(2\pi)^d} \bigl[ e^{\imath k x} u^+(k)
        + e^{-\imath k x} u^-(k) \bigr], \label{upm}
\end{align} 
where the operators
$
u^\pm(k) = u(\pm k) \theta (k_0)$  
are subjected to canonical commutation 
relations $$[u^+(k),u^-(k')]=\Delta(k,k').$$
In case of the scale-dependent fields, because of the presence of the scale argument in new fields 
$\phi_{a,\eta}(x)$, where $a$ and $\eta$ label the size and the shape 
of the region centered at $x$, the problem arises how to order the 
operators supported by different regions. This problem was solved 
in \cite{AltaiskyPEPAN2005,Altaisky2010PRD} on the base of the {\em region causality assumption} \cite{CC2005}. 
If two regions $(\Delta x,x)$ and $(\Delta y,y)$ do not intersect the standard 
time ordering procedure is applied. 
%(In light cone variables the $x_+$ argument is used instead of $x_0$). 
Alternatively, if one of the 
regions is {\em inside} another, see Fig.~\ref{tnew2:pic} the operator standing for the bigger 
region acts on vacuum first \cite{Altaisky2010PRD}.
%\begin{equation}
%T( A_{\Delta x}(x) B_{\Delta y}(y) ) = \begin{cases}
%A_{\Delta x}(x) B_{\Delta y}(y), & y_0 < x_0, \\
%\pm B_{\Delta y}(y) A_{\Delta x}(x), & x_0 < y_0, \\
%A_{\Delta x}(x) B_{\Delta y}(y), & \Delta x \subset \Delta y, \\
%\pm B_{\Delta y}(y) A_{\Delta x}(x), & \Delta y \subset \Delta x
%\end{cases}
%\label{tnew}
%\end{equation}
This causal ordering, drawn in Euclidean space, is 
presented in Fig.~\ref{tnew2:pic} below. The time ordering 
in Euclidean space, as an analytic continuation of time 
ordering in Minkowski space, have been already considered 
in \cite{EE1979}.
The diagram Fig.~\ref{tnew2:pic} shows space-like regions in Euclidean 
space. For Minkowski space corresponding diagrams can be obtained by 
analytic continuation of the Euclidean ball of imaginary radius $\imath\Delta$ 
into Minkowski space,  
%$$
% x^2 - t^2 \ge -\Delta^2, 
%$$ 
where we can restrict ourselves with forward light cone $t\ge0,|x|\le t$. 
The disjoint events in Minkowski space are shown in Fig.~\ref{mink1:pic}.
The correspondence to the other case of one Euclidean event inside 
another, shown in Fig.~\ref{tnew2:pic}b, looks more complex after 
analytic continuation to Minkowski space. The forward light-cone part 
of such intersection is shown in Fig.~\ref{mink2:pic}.

%%%%%%%%%%%%%%%%%%%%%%%%%%%%% Ins 2 %%%%%%%%%%%
We consider partial intersection of 
regions ($A \cap B =C, C\ne A, C\ne B, C\ne\emptyset$) as unphysical. For this reason 
corresponding ordering of operator-valued fields is not defined. Since a region is identified with 
a possibility of measurement, a simultaneous 
measurement of a part within and not within the 
parent entity is inconsistent. The "partial intersection" just implies that doing the functional 
integration one has to go to the finer scale, so 
that the regions do not intersect. The same 
happens in $p$-adic models of quantum gravity: 
two $p$-adic balls are either disjoint or one 
within another \cite{VVZ1994}.

Mathematically, when we make the functional 
measure of a Feynman integral 
into a a discrete product of wavelet fields on 
a lattice  $\mathcal{D}u_a(b)\to \prod_{j,k} d d^j_k$ we get rid of partial intersection, 
as it can be seen in the example of a binary 
tree, shown in the Table ~\ref{t1:tab}  below
\begin{table}
\begin{center}
\begin{tabular}{|c|c|c|c|}
\hline 
\multicolumn{2}{|c|}{$\underline{d^0_0}$} & \multicolumn{2}{c|}{$d^0_1$} \\
\hline
$d^1_{00}$ & $d^1_{01}$ & $\underline{d^1_{10}}$ & $d^1_{11}$ \\
\hline
\end{tabular}
\end{center}
\caption{Binary tree of operator-valued functions. Vertical direction corresponds to the scale variable. The causal sequence of the operator-valued functions shown in the table 
above is: $d^0_0, d^1_{00}, d^1_{01}, d^0_1, d^1_{10}, d^1_{11}$. As it is shown the underlined regions of different scales do not intersect}
\label{t1:tab}
\end{table} 
%%%%%%%%%%%%%%%%%%%%%%%%%%%%%%%%%%%%%%%%%%%%%%%

Phenomenologically, the principle 'the coarse acts first' is related to 
the definition of the measurement procedure, possibly generalized, 
where the state of a part can 
be measured/affected only after and relatively to the state of the whole. 
The similar reason underlies the restriction on the scales in internal 
loops by the minimal scales of all external lines of the Feynman diagram:
if we measure a quantum system from outside we cannot excite modes finer 
than the minimal available scale of measurement.
Thus the functional integration over the trajectories in the space of square-integrable functions $\cD \phi(x)$ is substituted by functional integration over all causal 
paths, or tubes of all different thickness, in the space of scale-dependent 
functions $\cD \phi_a(x)$. Referring the reader to the original works of 
\cite{Sorkin2003,CC2005} for the topological aspects of causal paths we ought mention 
that the Bogolioubov microcausality condition holds for causal paths in the 
same way as it holds for trajectories \cite{Altaisky2010PRD}. It is also easy 
to show, that if the domain $Y$ is inside the domain $X$ the corresponding Green 
function is not singular at coinciding arguments -- it is a projection 
from coarser scale to finer scale:
$$
G_0^{(2)}(a_1,a_2,b_1-b_2=0)= \int \dk{p}{4} 
\frac{\tilde g(a_1 p) \tilde g(-a_2 p)
}
{p^2+m^2
}e^{-\imath p \cdot 0},
$$
since $|\tilde g(p)|$ vanish at $p\to\infty$.

\subsection{Commutation relations}
In case of wavelet transform the positive- and negative-frequency 
part operators \eqref{upm} can be expressed using wavelet transform 
\begin{equation}
u_i^{\pm}(k) = \frac{1}{C_{g_i}} \int_{-\infty}^\infty d\eta 
\int_0^\infty \frac{da}{a} \tilde{g}_i (a M^{-1}(\eta) k) u^\pm_{ia\eta}(k),
\end{equation}
from where we can set \cite{Altaisky2010PRD}: 
\begin{align}\nonumber 
[u^+_{ia\eta}(k),u^-_{ja'\eta'}(k') ] &=& \delta_{ij} C_{g_i} a \delta(a-a') \delta (\eta-\eta')\times \\
&\times& [u^+(k),u^-(k')]
\end{align}
to ensure canonical commutation relations for $[u^+(k),u^-(k')]$.
%%%%%%%%%%%%%%%%%%%%%%%%%%%%%%%%%%%%%%%%%%%%%%%%%%%%%%%%%%%%%%

\subsection{The Dyson-Schwinger equation}
Ordering in scale argument results in the modification 
of the Dyson-Schwinger equation in the theory of scale-dependent 
functions. Let $G(x-y,a_x,a_y)$ be the bare field propagator, describing the propagation of the field from the 
region $(y,a_y)$ to the region $(x,a_x)$. Let 
$\cG(x-y,a_x,a_y)$ be the full propagator for the same regions. 
The Dyson-Schwinger equation relating the full propagator with the 
bare propagator is symbolically drawn in the diagram 
\begin{equation}
\feyn{\vertexlabel_{a_y} m  \vertexlabel_{a_x}} = 
\feyn{\vertexlabel_{a_y} f  \vertexlabel_{a_x}} + 
\feyn{\vertexlabel_{a_y} m  \vertexlabel_{a_1}   p \vertexlabel_{a_2} f \vertexlabel_{a_x}} \label{dyson:pic}
\end{equation}
The integral equation depicted in diagram \eqref{dyson:pic}
can be written as  
\begin{widetext}
\begin{equation}  
\cG(x-y,a_x,a_y) = G(x-y,a_x,a_y) + \int 
\frac{da_1}{a_1}\int \frac{da_2}{a_2} \int dx_1 dx_2 
G(x-x_2,a_x,a_2) \cP(x_2-x_1,a_2,a_1) \cG(x_1-y,a_1,a_y)
\label{dyson1x},
\end{equation}
\end{widetext}
where the full vertex $\feyn{p}=\cP(x_2-x_1,a_2,a_1)$ denotes the vacuum polarization operator if $G$ is the massless boson, or the self-energy diagram otherwise. The Fourier counterpart of the 
equation \eqref{dyson1x} can be written as 
\begin{widetext}
%\begin{equation}    
$$
\tilde\cG_{a_x,a_y}(p) =\tilde G_{a_x,a_y}(p) +  \int 
\frac{da_1}{a_1}\int \frac{da_2}{a_2} \tilde G_{a_x,a_2}(p) \tilde\cP_{a_2,a_1}(p) \tilde\cG_{a_1,a_y}(p).
$$
%\label{dyson1p}
%\end{equation}  
\end{widetext}

\subsection{Wavelet transform in Minkowski space}
The straightforward application of wavelet transform \eqref{dwtrd}, defined in Euclidean space $\R^d$, to the Minkowski 
space $M^4_1$ would be the analytic continuation of the results 
into the imaginary time $x_4 = \imath x_0$, making the Euclidean 
rotations into Lorentz boosts. The construction of such wavelets 
with respect to the representations of the Poincare group have 
been studied by several authors \cite{KlaStre91,ali1995coherent}. From physical point of 
view there exists a simple and an elegant way of making wavelet 
transform in Minkowski space. 

In quantum field theory problems related to relativistic particle collisions 
we can always change the coordinate frame to the co-moving frame  
of a relativistic projectile moving at utmost speed of light. 
Due to the Lorentz contraction of the projectile the longitudinal and the transversal degrees of freedom behave essentially different 
in such system. Without loss of generality we can always assume 
the projectile to move along the $z$ axis. 

The Lorentz contraction, i.e., the hyperbolic rotation in ($t,z$) plane 
is determined by the hyperbolic rotation angle -- the rapidity $\eta$. 
The rotations in the transverse plane are not affected by the Lorentz 
contraction and are determined by the $SO(2)$ rotation angle $\phi$. 
If the problem is axially symmetric, the latter can be dropped.

Therefore it is convenient to change from the 
space-time coordinates $(t,x,y,z)$ to the {\em light-cone coordinates} $(x_+,x_-,x,y)$:
\begin{equation}
x_\pm = \frac{t \pm z}{\sqrt{2}},\quad \vx_\perp = (x,y).
\label{lcc}
\end{equation}
This is the so-called infinite momentum frame. The IMF is not 
a Lorentzian system, but a limit of that at $v\to c$. 
%The IMF very 
% much fits the problems of relativistic heavy ion collisions 
% with projectiles moving at utmost speed of light. 
The advantage of the coordinates \eqref{lcc} for the calculations, say in QED or QCD, 
is significant simplification of the vacuum structure \cite{ChangMa1969,KS1970}.
The metrics in the light-cone coordinates becomes 
$$
g_{\mu\nu} = \begin{pmatrix}
0 & 1 & 0 & 0 \cr 
1 & 0 & 0 & 0 \cr 
0 & 0 &-1 & 0 \cr
0 & 0 & 0 & -1
\end{pmatrix}.
$$
The rotation matrix has a block-diagonal form 
$$
M(\eta,\phi) = \begin{pmatrix}
e^{\eta} & 0 & 0 & 0 \cr
0 & e^{-\eta} & 0 & 0 \cr
0 & 0 & \cos\phi & \sin\phi \cr
0 & 0 &-\sin\phi & \cos\phi
\end{pmatrix}, 
$$
%\label{lcrot}
%\end{equation}
so that $M^{-1}(\eta,\phi)=M(-\eta,-\phi)$.

We can define the wavelet transform in light-cone coordinates 
in the same way as in Euclidean space using the representation 
of the affine group 
$$
x' = a M(\eta,\phi) x + b, $$
defined in $L^1$ norm as 
%\begin{equation}
$$
U(a,b,\eta,\phi)u(x) = \frac{1}{a^4}u\left(M^{-1}(\eta,\phi)\frac{x-b}{a} \right). 
$$
%\label{mrep}
%\end{equation}
we have the definition of wavelet coefficients of a function $f(x)$ 
with respect to the basic wavelet $g$ as follows
\begin{align}\nonumber 
W_{a,b,\eta,\phi}[f] &=& \int dx_+ dx_- d^2 \vx_\perp 
\frac{1}{a^4} \times \\ &\times& \overline{ g\left(M^{-1}(\eta,\phi)\frac{x-b}{a} \right)} f(x_+,x_-,\vx_\perp). 
\label{dwtm}
\end{align}
In contrast to wavelet transform in Euclidean space, where the 
basic wavelet $g$ can be defined globally on $\R^d$, the basic 
wavelet in Minkowski space is to be defined separately in four 
domains impassible by Lorentz rotations:
\begin{align*} 
A_1:  k_+ >0, k_-<0; & 
A_2:  k_+ <0, k_- >0;\\
A_3:  k_+ >0, k_- >0;& 
A_4:  k_+ <0, k_-<0 ,
\end{align*}
where $k$ is wave vector, $k_\pm = \frac{\omega \pm k_z}{\sqrt{2}}$. 
Whence we have four separate wavelets in these four domains \cite{PG2011}:
\begin{equation}
g_i(x) = \int_{A_i} e^{\imath k x} \tilde{g}(k) \dk{k}{4}.
\end{equation}
We assert the following definition of the Fourier transform 
in light cone coordinates:  
\begin{align*}\nonumber 
f(x_+,x_-,\vx_\perp) = \int e^{\imath k_- x_+ + \imath k_+ x_- -\imath \vk_\perp \vx_\perp} \tilde{f} (k_-, k_+,\vk_\perp) \times \\ \frac{dk_+dk_-d^2\vk_\perp}{(2\pi)^4}. 
%\label{ftlc}
\end{align*}
Substituting the Fourier images into the definition \eqref{dwtm} 
we get 
\begin{align}\nonumber 
W_{ab\eta\phi}^i = \int_{A_i} e^{\imath k_- b_+ + \imath k_+ b_- -\imath \vk_\perp \vb_\perp} \tilde{f} (k_-, k_+,\vk_\perp) \\ \overline{\tilde{g}}(a e^\eta k_-, a e^{-\eta} k_+, a R^{-1}(\phi) \vk_\perp) \frac{dk_+dk_-d^2\vk_\perp}{(2\pi)^4}.
\end{align}
In Fourier space the relation between Fourier coefficients and wavelet 
coefficients is therefore the same as in $\R^d$:
$$
\tilde{W}_{a\eta\phi}(k) = \tilde{f}(k) \overline{\tilde{g}}(aM^{-1}(\eta,\phi)k).
$$
Similarly, the reconstruction formula is \cite{AK2013iv}:
\begin{widetext} 
\begin{align*}\nonumber 
f(x) &=&  \sum_{i=1}^4 \frac{1}{C_{g_i}} \int_{-\infty}^\infty d\eta 
\int_0^{2\pi} d\phi
\int_0^\infty \frac{da}{a} \int_{M^4_1} db_+ db_- d^2\vb_\perp  
\frac{1}{a^4} g_i \left( M^{-1}(\eta)\frac{\xi-b}{a} \right) W^i_{ab\eta\phi} \\
&=& \sum_{i=1}^4 \frac{1}{C_{g_i}} \int_{-\infty}^\infty d\eta  
\int_0^{2\pi} d\phi 
\int_0^\infty \frac{da}{a}
 \int_{A_i} \frac{dk_+dk_-d^2\vk_\perp}{(2\pi)^4}
e^{\imath k_- x_+ + \imath k_+ x_- -\imath \vk_\perp \vx_\perp}
\tilde{W}_{a\eta\phi}(k) \tilde{g}(ak_- e^\eta, a k_+ e^{-\eta},aR^{-1}(\phi)\vk_\perp)  
%\label{iwtm}
\end{align*}
\end{widetext}
If the problem is axially symmetric the azimuthal part of integration 
($\phi$) can be dropped. It is also convenient to use logarithmic scale $a = e^u$ to study different rapidity domains.
%%%%%%%%%%%%%%%%%%%%%%%%%%%%%%%%%%%%%%%%%%%%%%%%%%%%%%%%%%%%%%

\subsection{Choice of the basic wavelet}
The choice of the basis of wavelet decomposition is 
always a subtle question, specially in quantum field 
theory. (The best choice, as it was already emphasized in   \cite{Altaisky2010PRD}, 
would be the apparatus function of a classical measuring 
device interacting with quantum system.) Some basis is always tacitly assumed. Even 
describing the massless photons, which are not localized 
anywhere, by plane waves, the possibility of photon registration by photomultiplier implies its interaction 
with electron, and hence some scale and some localization.

If the continuous wavelet transform is used in place of the 
Fourier transform the choice of the basic function is 
constrained by the feasibility of the analytical integration 
in Feynman diagrams on one hand, and by the possibility 
to understand this basic function as a localized (quasi) particle. The latter 
being claimed by some authors to be important for 
Minkowski space \cite{PS2007}, seems questionable 
for Euclidean space calculations. If the wavelet transform 
is performed on a lattice there is a bias that only the 
similarity properties are important, rather than the shape 
of wavelet \cite{Battle1999,BP2013}. The question whether or not 
the basic wavelet should satisfy some equation of motion 
is still not clear.
%%%%%%%%%%%%%%%%%%%%%%% Ins 3 %%%%%%%%%%%
We are also not aware of the effect of the 
discrete symmetries of the basic wavelet.
%%%%%%%%%%%%%%%%%%%%%%%%%%%%%%%%%%%%%%%%%

\begin{figure}[ht]
\centering \includegraphics[width=4cm]{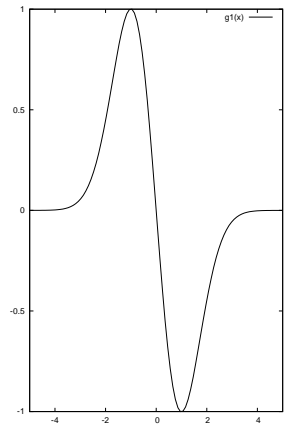}
\caption{Graph of $g_1$ wavelet: $g_1(x)=-x e^{-x^2/2}$}
\label{g1:pic}
\end{figure}

To justify our choice of the derivatives of the Gaussian 
as the basic wavelets, we present the following heuristic 
consideration, inferred from the coherent states theory 
\cite{ali1995coherent}. 
Let us introduce a localized wave packet in Fourier space  
\begin{equation}
\tilde g(t,k) = e^{-\imath t k - k^2/2}. 
\label{gpcc}
\end{equation}
If the wave packet  is considered in Minkowski 
space, then $k^2=0$ can be assumed for the photon and 
the whole equation \eqref{gpcc} turns to be a plane wave.
Otherwise it is a localized wave. If $t$ is time the packet 
\eqref{gpcc} is a gaussian wave packet at initial time $t\!=\!0$.
At finite but small instants of time the wavepacket 
can be approximated by its Taylor expansion 
$$
\tilde g(t,k) = \tilde g_0(k) + \frac{t}{1!} \tilde g_1(k) 
              + \frac{t^2}{2!} \tilde g_2(k) + O(t^3), 
$$
where the expansion coefficients  
$$\tilde g_n(k) = \left. \frac{d^n}{dt^n}\tilde g(t,k) \right|_{t=0}$$
are responsible for the shape of the packet at the time scales 
at which 1, 2 or $n$ excitations are significant. Clearly 
$g_n(x)$ are the excitations of the harmonic oscillator, with 
$g_1$ being the first excitation, see Fig.~\ref{g1:pic}.

\section{Gauge theories \label{gau:sec}}

\subsection{QED}
Quantum electrodynamics is the simplest case of gauge theory. 
The local $U(1)$ invariance of the fermion field 
$$\psi(x) \to e^{-\imath e \Lambda(x)}\psi(x)$$
 is 
accompanied by the gradient invariance of the vector field $A_\mu(x)$, 
the electromagnetic field,
\begin{equation}
A_\mu(x) \to A_\mu(x) + \partial_\mu \Lambda(x),
\label{gau1}
\end{equation}
to keep the total action $S(\bar\psi,\psi,A)=\int L d^4x$ invariant under 
the local $U(1)$ transform generated by $\Lambda(x)$. 
The interaction of the charged fermion field $\psi$ with 
electromagnetic field $A_\mu$ is introduced by substitution of 
ordinary derivatives $\d_\mu$ to covariant derivatives 
$$
D_\mu = \d_\mu + \imath e A_\mu(x),
$$   
with $e$ being the charge of the fermion.

The Lagrangian of QED has the (Euclidean) form 
\begin{align}\nonumber
L &=& \bar\psi(x)(\slashed{D}+ \imath m)\psi(x) + \frac{1}{4} F_{\mu\nu}F^{\mu\nu}+& \underbrace{\frac{1}{2\alpha} (\d_\mu A_\mu)^2}_{\hbox{gauge fixing}}, \\
& &\label{gau1l}\hbox{with\ }  F_{\mu\nu}=\d_\mu A_\nu - \d_\nu A_\mu
\end{align}
being the field strength tensor of the electromagnetic field $A$, and 
$\gamma_\mu$ being the Dirac $\gamma$-matrices.

The straightforward application of the Feynman integral to the gauge 
theory with the Lagrangian \eqref{gau1l} would be inefficient for 
the integration over the field $A(x)$ would contain an infinite  
set of physically equivalent field configurations. For this purpose the gauge fixing, which restricts the integration only to gauge-nonequivalent 
field configurations was introduced by Faddeev and Popov \cite{FP1967}. 

Quantum electrodynamics is the most firmly established  and most 
verified field theory model in physics of elementary particles. 
The probability amplitude of scattering obtained at a tree-level 
are in fairly good agreement with classical scattering theory. 
Starting from one loop level the Feynman integrals are formally 
divergent, and the physical results are derived using the renormalization 
invariance of QED. The most accurate tests for the renormalized 
calculations of the electron-photon interaction are the Lamb shift 
of the Hydrogen-like ion energy levels and the anomalous magnetic 
momentum of the electron \cite{KL1949,Salpeter1952,Salpeter1953,EY1965}.

In one-loop approximation the radiation corrections in QED come from three primitive Feynman graphs: fermion self-energy 
$\Sigma(p)$, vacuum polarization operator $\Pi_{\mu\nu}(p)$,
and the vertex function $\Gamma_{\rho}(p,q)$.
%shown in Fig.~\ref{l1qed:pic} below. 
%\begin{figure}[ht]
%\centering \includegraphics[width=.2\textheight]{l1qed.eps}
%\caption{One-loop radiative corrections in QED: a) electron self-energy diagram; b) vacuum polarization diagram; c) vertex function}
%\label{l1qed:pic}
%\end{figure}
In Euclidean space the equations for the above three graphs %~\ref{l1qed:pic}a,b,c 
have the form:\\

Electron self-energy 
\begin{equation}
\Sigma(p) = -e^2 \int \dk{q}{4} \gamma_\mu \frac{-\imath}{\slashed{p}-\slashed{q}+m} \gamma_\nu \frac{\delta_{\mu\nu}}{q^2} \label{S2e} 
\end{equation}
gives the corrections to the bare electron mass $m_0$ related 
to irradiation of virtual photons.

Vacuum polarization diagram
\begin{equation}
\Pi_{\mu\nu}(p)  = -e^2 \int \dk{q}{4} {\rm Sp}[
\gamma_\mu \frac{1}{\slashed{p}+\slashed{q}+m}\gamma_\nu \frac{1}{\slashed{q}+m}
] \label{G2p}
\end{equation}
could be expected to give the nonzero corrections to the photon 
mass, but due to gauge invariance the photon mass remains zero, 
instead the one loop contribution \eqref{G2p} renormalizes the 
electron charge at small distances, therefore modifies the Coulomb 
interaction by screening the bare electron charge $e_0$ by 
virtual electron-positron pairs polarizing the vacuum at small 
distances. This diagram contributes to the Lamb shift of the 
atom energy levels.
 
One-loop vertex function
\begin{equation} 
\Gamma_\rho(p,q) = -\imath e^3 \int \dk{f}{4}\gamma_\tau 
\frac{1}{\slashed{p}+\slashed{f}+m} \gamma_\rho 
\frac{1}{\slashed{f}+\slashed{q}+m} \gamma_\sigma 
\frac{\delta_{\tau\sigma}}{f^2} \label{G1e}
\end{equation}
determines the anomalous magnetic moment of the electron.

All three integrals (\ref{S2e},\ref{G2p},\ref{G1e}) are  divergent. 
Their evaluation involves regularization procedures. The most common 
is the dimensional regularization with all integrals taken in formal 
$d=2\omega$ dimension with physical value $\omega=2$.  In 
this way the divergences come as poles in $\epsilon=2-\omega$,
see \eg \cite{HooftVeltman1972,Bsh1980,Ramond1989}. 

%At the vicinity of physical dimension $d\to4,\epsilon\to0$, for the 
%electron self-energy, vacuum polarization, and the vortex shown in 
%Fig.~\ref{l1qed:pic} we have, 
%respectively:
%\begin{align*}\nonumber 
%\Sigma &=& -\imath \left(e\mu^{2-\omega}\right)^2 
%\int \dk{q}{2\omega} 
%\frac{
%\gamma_\mu (\slashed{p}-\slashed{q}-m)\gamma_\nu \delta_{\mu\nu} 
%}
%{[(p-q)^2+m^2]q^2}  
%\\
%&=&-\frac{\imath e^2}{16\pi^2\epsilon}(\slashed{p}+4m)
% + \frac{\imath e^2}{8\pi^2}\Bigl[ \frac{\slashed{p}(1+\gamma)}{2} + 
% m(1+2\gamma) \\ \nonumber  
% &+& \int_0^1	dx 
%\left[ \slashed{p}(1-x) + 2m \right] 
%\ln \frac{p^2 x(1-x)+m^2 x}{4\pi\mu^2}
%\Bigr] \\ & &+ O(\epsilon) \\
%\Pi_{\mu\nu} &=& -\left(e\mu^{2-\omega}\right)^2  
%\int \dk{q}{2\omega} 
%\frac{
%{\rm Sp}\gamma_\mu [\slashed{p}+\slashed{q}-m]\gamma_\nu [\slashed{q}-m]} 
%{[q^2+ m^2][(p+q)^2+m^2]}  
%\\
%&=& \frac{e^2}{2\pi^2}(p_\mu p_\nu - p^2 \delta_{\mu\nu})\Bigl[ 
%\frac{1}{6\epsilon} - \frac{\gamma}{6} - \\
%&-& \int_0^1 dx x (1-x) \ln \frac{m^2+p^2 x(1-x)}{2\pi\mu^2} 
%+ O(\epsilon) \\
%\Gamma_k &=& \Gamma_k^{(1)}(p,q) + \Gamma_k^{(2)}(p,q)
%\end{align*}
%where $\mu$ is an arbitrary scale parameter of the mass dimension, $\gamma$ is Euler-Masceronni constant, $x$ is Feynman parameter.
%The equations for the vertex functions $\Gamma_k^{(i=1,2)}$ are given in Appendix. 

\subsection{One loop corrections in wavelet-based theory}
The evaluation of Feynman diagrams with fermions and gauge fields 
in wavelet-based Euclidean theory is similar to that of scalar theory \eqref{I4}. 
The evaluation of the one-loop radiative corrections for 
the scale-dependent fields give finite results by construction 
with no regularization procedure required.   

\begin{figure}[ht]
\centering \includegraphics[width=4cm]{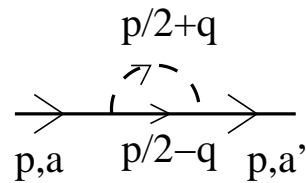}
\caption{Electron self-energy diagram}
\label{esea:pic}
\end{figure}

\begin{figure}[ht]
\centering \includegraphics[width=4cm]{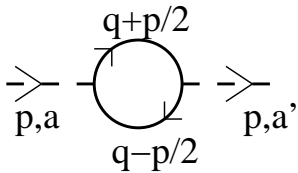}
\caption{Vacuum polarization diagram in (Euclidean) scale-dependent QED}
\label{vpd:pic}
\end{figure}

\paragraph{Electron self-energy.} 
For the scale components of the electron self-energy 
diagram, we get
\begin{equation}
\frac{\Sigma^{(A)}(p)}{\tilde g(a p) \tilde g(-a' p)} = -\imath e^2 \int \dk{q}{4}  
\frac{F_A(p,q)
\gamma_\mu 
 \left[\frac{\slashed{p}}{2}-\slashed{q}-m \right] \gamma_\mu
 }
 {\left[
\left(\frac{p}{2}-q \right)^2+m^2\right]
 \left[\frac{p}{2}+q \right]^2
 },
\end{equation}
where $A$ is the minimal scale of two external lines shown in Fig.~\ref{esea:pic}: $A= \min(a,a')$. The regularizing function $F_A(p,q)$ is the 
result of integration over the scales of two internal lines.
For the isotropic basic wavelet $g$ the regularizing 
function is given by \eqref{cutf1}:
\begin{equation}
F_A(p,q) = f^2(A(p/2-q)) f^2(A(p/2+q)). \label{FA}
\end{equation}
Introducing the dimensionless variable $\vy=\vq/|\vp|$, after straightforward algebra, we can perform the integration in Euclidean space: 
\begin{align} \label{sa1:eq}
\frac{\Sigma^{(A)}(p)}{\tilde g (ap) \tilde g(-a'p)} = -\imath e^2 \int  \dk{y}{4}F_A(p,|p|y)\times\\
\nonumber \times  \frac{
 \slashed{p}+4m -2 |p|\slashed{y}
}{
\left[ 
y^2 + \frac{1}{4} - y \cos\theta - \frac{m^2}{p^2} 
\right]
\left[ 
y^2 + \frac{1}{4} + y \cos\theta 
\right]
}.
\end{align} 
where $\theta$ is the Euclidean angle between the $p$ and the $q$ 
directions. 
%For the isotropic case 
%$
%d^4 y = 4\pi \sin^2 \theta d\theta y^3 dy.
%$
In high energy limit, $p^2 \gg 4 m^2$, the contribution of last term in the numerator 
of \eqref{sa1:eq} vanishes for the symmetry, and the equation 
\eqref{sa1:eq} can be easily integrated in angle variable \eqref{Ik}:
\begin{align*} 
%\label{sa2}
\frac{\Sigma^{(A)}(p)}{\tilde g(ap) \tilde g(-a'p) } &=&-\frac{\imath e^2}{4\pi^2} R_1(p)(\slashed{p}+4m) \qquad \hbox{where: }\\
\nonumber R_1(p) &=& \int_0^\infty 
dy y F_A(p,|p|y) \left[1 - \sqrt{1-\frac{1}{\beta^2(y)}} \right], \\
\nonumber & &\beta(y) = y + \frac{1}{4y}.
\end{align*}
The integral $R_1(p)$ is finite for any wavelet cutoff function 
\eqref{cutf1}.
For the $g_1$ wavelet we get 
\begin{align*}
R_1(p) &=& e^{-A^2 p^2} \int_0^\infty dy y e^{-4A^2p^2 y^2}
\left[1-\sqrt{1-\frac{1}{\beta^2}} \right].
\end{align*}
After a simple algebra this gives 
\begin{align} \nonumber
R_1(p) &=& \frac{1}{8A^2p^2} 
\big(2 A^2p^2 \Ei_1(A^2p^2)- 4 A^2 p^2 \Ei_1(2A^2 p^2) \\
& -&e^{-A^2p^2}   +2 e^{-2A^2p^2} 
\big)
\end{align}
\paragraph{Vacuum polarization diagram}
The vacuum polarization diagram in quantum electrodynamics 
of scale-dependent fields is obtained by integration over 
the scale variables of the fermion loop shown in Fig.~\ref{vpd:pic}:
\begin{align}\nonumber 
\frac{\Pi_{\mu\nu}^{(A)}(p)}{\tilde g (ap) \tilde g(-a'p)}&=& -e^2 \int \dk{q}{4} F_A(p,q) \times \\
\nonumber &\times& \frac{{\rm Sp} (\gamma_\mu (\slashed{q}+ \slashed{p}/2  -m)\gamma_\nu (\slashed{q} -\slashed{p}/2 - m))}{\left[(q+p/2)^2+m^2\right]\left[(q-p/2)^2+m^2\right]} \\
\nonumber &=& - 4e^2 \int \dk{q}{4} F_A(p,q) \times  \\ 
 &\times& \frac{2q_\mu q_\nu - \frac{1}{2} p_\mu p_\nu + 
\delta_{\mu\nu}(\frac{p^2}{4}-q^2-m^2)}{\left[(q+\frac{p}{2})^2+m^2\right]\left[(q-\frac{p}{2})^2+m^2\right]}.
\label{padef}
\end{align}
Similarly to previous diagram we use the dimensionless variable $y$ and 
integrate over the angle variable.
 %%%%%%%%%%%%%%%%%%%%%%%%%%%%%%%%%%%%%%%%%%%%%%%%%%%%%%%%%%%%%%
The momentum integration in equation \eqref{padef} is straightforward: having expressed all momenta in units 
of electron mass $m$, we express the loop momentum in terms of the photon momentum  
and perform the integration over the angle variable:    
\begin{eqnarray*}
\nonumber \frac{\Pi_{\mu\nu}^{(A)}}{\tilde{g}(ap)\tilde{g}(-a'p)} = - \frac{e^2}{\pi^3} (m^2p^2) \int_0^\infty 
dy y F_A(m p,m p y) \times \\ 
\times \int_0^\pi d\theta \sin^{2}\theta 
\frac{2y_\mu y_\nu - \frac{1}{2} \frac{p_\mu p_\nu}{p^2} + 
\delta_{\mu\nu}(\frac{1}{4}-y^2-\frac{1}{p^2})}
{\left[\frac{\frac{1}{4}+y^2+\frac{1}{p^2}}{y}+\cos\theta\right]
\left[\frac{\frac{1}{4}+y^2+\frac{1}{p^2}}{y}-\cos\theta\right]
}, 
\end{eqnarray*} 
where $p$ is dimensionless, \ie is expressed in units of $m$. Introducing the notation 
$\beta(y)\equiv \frac{\frac{1}{4}+y^2+\frac{1}{p^2}}{y}$ and using the substitution 
$
y_\mu y_\nu \to A y^2 \delta_{\mu\nu} + B y^2 \frac{p_\mu p_\nu}{p^2},$
we get 
\begin{eqnarray*}
\nonumber \frac{\Pi_{\mu\nu}^{(A)}}{\tilde{g}(ap)\tilde{g}(-a'p)} = - \frac{e^2}{\pi^3} (m^2p^2) \int_0^\infty 
dy y F_A(mp,mpy) \times \\
\times \int_0^\pi d\theta \sin^{2}\theta  
\frac{ 
\delta_{\mu\nu}( (2A-1)y^2 + \frac{1}{4}-\frac{1}{p^2}) + \frac{p_\mu p_\nu}{p^2} (2By^2-\frac{1}{2})
}
{\beta^2(y)-\cos^2\theta},
\end{eqnarray*}
where $A$ and $B$ depend only on the modulus of $y$, but not on the direction, and can be expressed 
in terms of angle integrals \eqref{Ik}.

Finally, writing the polarization operator as a sum of transversal and longitudinal 
parts, we have the equations 
%\begin{widetext}
\begin{align} \nonumber 
\frac{\Pi_{\mu\nu}^{(A)}(p)}{\tilde{g}(ap)\tilde{g}(-a'p)} &\equiv& \delta_{\mu\nu} \pi^{(A)}_T + \frac{p_\mu p_\nu}{p^2} \pi^{(A)}_L \\ \nonumber  
&=& \left(\delta_{\mu\nu}-\frac{p_\mu p_\nu}{p^2} \right)\pi^{(A)}_T + X^{(A)} \frac{p_\mu p_\nu}{p^2} ,
%\label{pia} 
\\ 
\pi^{(A)}_T &=& -\frac{e^2}{3\pi^2} m^2p^2 \int_0^\infty dy y F_A(mp,mpy)
\Bigl[y^2 + \label{piat} \\ \nonumber
&+& \left(1-\sqrt{\frac{\frac{1}{16}+y^4+\frac{1}{p^4}-\frac{y^2}{2}+\frac{1}{2p^2}+\frac{2y^2}{p^2}}
{\left(\frac{1}{4} + y^2 + \frac{1}{p^2}\right)^2}} \right) \\
\nonumber &\times& \left(\frac{5}{8}-\frac{4}{p^2}-\frac{2}{p^4}-2y^2\left(1+\frac{2}{p^2}\right)-2y^4\right)\Bigr] 
\\ \nonumber 
\pi^{(A)}_L &=& -\frac{e^2}{3\pi^2} m^2p^2 \int_0^\infty dy y F_A(mp,mpy)
\Bigl[-4y^2 + 
%\label{pial} 
\\ \nonumber
&+& \left(1-\sqrt{\frac{\frac{1}{16}+y^4+\frac{1}{p^4}-\frac{y^2}{2}+\frac{1}{2p^2}+\frac{2y^2}{p^2}}
{\left(\frac{1}{4} + y^2 + \frac{1}{p^2}\right)^2}} \right) \\
\nonumber &\times& \left(8y^4+2y^2\left(1+\frac{8}{p^2}\right)+\frac{4}{p^2}+\frac{8}{p^4}-1\right)\Bigr],
\end{align}
where 
\begin{align*}
X^{(A)} &=&\pi_L^{(A)} + \pi_T^{(A)} = \frac{e^2 m^2 p^2}{\pi^2} \int_0^\infty dy y F_A(mp,mpy)  \\
&\times& \Bigl[y^2 - 
\left(1-\sqrt{\frac{\frac{1}{16}+y^4+\frac{1}{p^4}-\frac{y^2}{2}+\frac{1}{2p^2}+\frac{2y^2}{p^2}}
{\left(\frac{1}{4} + y^2 + \frac{1}{p^2}\right)^2}} \right) \\
\nonumber &\times& \left(2y^4+4 \frac{y^2}{p^2}+\frac{2}{p^4}-\frac{1}{8}\right)\Bigr].
\end{align*}
%\end{widetext}
The integrals above are finite and can be easily evaluated in large momenta limit, $p^2\gg 4$, introducing the dimensionless scale $a=Am$.

As an example we can evaluate the vacuum polarization operator 
for $g_1$ wavelet. For $g_1$ wavelet the regularizing function 
$$
F_A(p,q) = \exp\big(-A^2p^2 - 4 A^2 q^2\big).$$
Hence for large  $p^2 \gg 4$ the integral 
\eqref{piat} can be evaluated by substitution $y^2 = t$ \cite{AltSIGMA07}:
\begin{align*}
\pi^{(A)}_T 
%&=& -\frac{e^2}{6\pi^2} m^2p^2 \int_0^\infty  dt \exp\left(-A^2m^2p^2(1+4t)\right) \\
%&\times& \left[t + \left(1-\sqrt\frac{(\frac{1}{4}-t)^2}{(\frac{1}{4}+t)^2 }\right) 
%\left(\frac{5}{8}-2t -2t^2\right) \right] \\
 &=& -\frac{e^2}{6\pi^2} m^2 p^2
\Bigl\{
\frac{e^{-a^2p^2}}{8a^6p^6}\big(4a^4p^4-a^2p^2-1\big) 
+\frac{e^{-2a^2p^2}}{8a^6p^6}  \\
&\times& \big(1-4a^4p^4+2a^2p^2\big) 
-\frac{1}{2}\Ei_1\big(a^2p^2\big) + \Ei_1\big(2a^2p^2\big) %\label{pt-inf}
\Bigr\}. 
\end{align*}
Similarly, the longtitudinal term $X^A$ evaluated with $g_1$ 
wavelet in the limit $p^2 \gg 4$ is equal to 
\begin{equation}
X^A = \frac{e^2 m^2 p^2}{16\pi^2} \frac{e^{-a^2 p^2}(a^2 p^2 -1 + e^{-a^2 p^2})}{a^6p^6} \label{xap}
\end{equation}
In the limit of small scales $ap \ll 1$ the equation \eqref{xap} does not depend on $p$: $X^A \propto \frac{1}{a^2}$. Therefore the whole 
equation \eqref{piat} is similar to the result obtained 
by Pauli-Villars regularization of the vacuum polarization 
%\begin{equation}
$$
\Pi^M_{\mu\nu}(p) = c M^2 \delta_{\mu\nu} +(p^2\delta_{\mu\nu}-p_\mu p_\nu)
F\left(\frac{p^2}{4m^2},\frac{m}{M} \right),
$$
%\label{pve},
%\end{equation}
where $M\to\infty$ is a regularizing mass \cite{Bsh1980}. 
The gauge invariance is restored if the multiscale diagram
\eqref{padef} is integrated over all scales. In this limit 
the theory can be subjected to dimensional regularization 
\cite{Altaisky2010PRD}.
%%%%%%%%%%%%%%%%%%%%%%%%%%%%%%% 
\paragraph{Vertex function.} The one-loop contribution to the 
QED vertex function for the theory with scale-dependent matter fields 
is shown in Fig.~\ref{rc1:pic} below.
\begin{figure}[ht]
\centering \includegraphics[width=.4\textwidth]{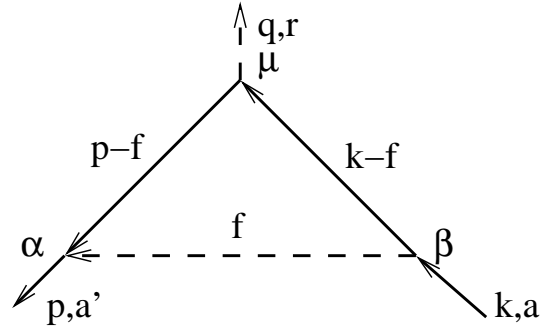}
\caption{One-loop vertex function in scale-dependent QED} \label{rc1:pic}
\end{figure}
The equation, which corresponds to the 
vertex diagram Fig.~\eqref{rc1:pic} can be casted in the form 
\begin{align*}
-\imath e \frac{\Gamma_{\mu,r}^{(A)}}{\tilde{g}(-pa') \tilde{g}(-qr) \tilde{g}(ka)}  = (-\imath e)^3 
\int  \dk{l}{4} \gamma_\alpha G(p-f) \gamma_\mu \times \\
\times G(k-f) \gamma_\beta D_{\alpha\beta} F_A(p-f) F_A(k-f) F_A(f). 
\end{align*}
The explicit substitution with photon propagator taken in Feynman gauge 
gives 
\begin{align}\nonumber 
\imath e \frac{\Gamma_{\mu,r}^{(A)}}{\tilde{g}(-pa') \tilde{g}(-qr) \tilde{g}(ka)}  = (-\imath e)^3 
\int  \dk{f}{4} \gamma_\alpha 
\frac{\slashed{p}-\slashed{f}-m}{(p-f)^2+m^2}  \\
\times \gamma_\mu   \frac{\slashed{k}-\slashed{f}-m}{(k-f)^2+m^2}    \gamma_\alpha \frac{1}{f^2} F_A(p-f) F_A(k-f) F_A(f) \label{l1v}
\end{align}
Representing the numerator of the latter equation in the form 
\begin{align*}
A_\mu &=& \gamma_\alpha (\slashed{p}-\slashed{f})\gamma_\mu  (\slashed{k}-\slashed{f})  \gamma_\alpha - \\
&-&m \left[ 
(\slashed{p}-\slashed{f})\gamma_\mu + \gamma_\mu (\slashed{k}-\slashed{f}) 
\right] + 2 m^2 \gamma_\mu, 
\end{align*}
it can be seen that the right-hand side of the equation \eqref{l1v} 
can be represented as a linear combination of three finite integrals 
($J^{(0)},J^{(1)}_\mu,J^{(2)}_{\mu\nu}$) presented in Appendix \ref{gfi:app}, in analog  to their divergent counterparts in Minkowski space \cite{KRP1995}. After some algebra the vertex \eqref{l1v} turns 
to be 
\begin{align}\nonumber 
\frac{\Gamma_{\mu,r}^{(A)}}{\tilde{g}(-pa') \tilde{g}(-qr) \tilde{g}(ka)}  &=& e^2 \gamma_\alpha [(\slashed{p}\gamma_\mu \slashed{k} - m \slashed{p}\gamma_\mu -m \gamma_{\mu} \slashed{k}) J^{(0)} \\
\nonumber &-& (\gamma_\nu \gamma_\mu \slashed{k} 
+ \slashed{p} \gamma_\mu \gamma_\nu + 2m \delta_{\mu\nu}) J^{(1)}_\nu \\
&+& \gamma_\lambda \gamma_\mu \gamma_\nu J^{(2)}_{\lambda\nu}]\gamma_\alpha.
\end{align}

\subsection{Ward-Takahashi identities}
The Ward-Takahashi identity in spinor electrodynamics 
relates the vertex function to the difference of fermion propagators:
\begin{equation}
q_\mu \Gamma_\mu(p,q,p+q) = G^{-1}(p+q) - G^{-1}(p),
\label{wt1}
\end{equation}
where $G(p)$ is the complete fermion propagator. The identity 
\eqref{wt1} is a helpful constraint which ensures the gauge invariance 
of the renormalized QED in any order of perturbation theory 
\cite{Ward1950,Takahashi1957}. The constraint \eqref{wt1} makes 
the perturbation expansion gauge invariant at the presence of the 
gauge fixing terms in the QED generating functional. 

In wavelet-based theory, where the fields explicitly depend on scale 
the divergence do not appear in the Feynman diagrams, but the evaluation 
of integrals in internal lines with the integration scales constrained 
by the minimal scale of external lines may spoil the gauge invariance 
of the complete propagator. To prevent this the Ward-Takahashi identities are required.

At the absence of gauge-fixing terms in the Lagrangian \eqref{gau1l}, 
the generating functional 
\begin{equation}
e^{-Z[J,\bar\chi,\chi]}= \int \cD A \cD \bar\psi \cD \psi 
e^{-\int d^4 x (L(\psi,\bar\psi,A) + JA + \imath \bar{\chi}\psi +\imath \bar{\psi}\chi)}, \label{qed-gf} 
\end{equation}
with $L(\psi,\bar\psi,A)$ given by the equation \eqref{gau1l}, would 
be invariant under the gauge transformations \eqref{gau1} if no source 
term $-\int d^4x (J_\mu A_\mu + \imath \bar{\chi}\psi +\imath \bar{\psi}\chi)$ is present.

In the framework of scale-dependent functions the gauge field $A_\mu(x)$ is expressed in terms of its wavelet coefficients $A_{\mu a}(b)$:
$$
A_\mu(x) = \frac{1}{C_g} \int_{\R_+ \times \R^d} \frac{1}{a^d}
g \left(\frac{x-b}{a} \right) A_{\mu a}(b) \frac{da d^d b}{a}
$$
(with the angular part of wavelet transform \eqref{iwt} dropped 
for simplicity). In view of linearity of wavelet transform we may 
infer the gauge transform of the scale components to have 
the form 
%\begin{equation}
$$
A_{\mu a}'(x) = A_{\mu a}(x) + \frac{\d \Lambda_a(x)}{\d x_\mu},
$$
%\label{gau1a}
%\end{equation}
where 
$$
\Lambda_a(x) = \int_{\R^d} \frac{1}{a^d} \bar{g} \left(
\frac{y-x}{a}
\right) \Lambda(y) d^dy
$$
is the scale component of the gauge function \eqref{gau1}. That is the 
gauge transform of the abelian gauge field $A_{\mu a}(x)$ is  
a projection of the (no-scale) gauge field $A_\mu(x)$ onto the 
space of resolution $a$.

Since the free Lagrangian of QED is gauge-invariant by construction, 
the derivative of the Ward-Takahashi identities turns into evaluation of the functional 
overage of the variation of source and gauge fixing terms under 
inifinitesimal gauge transform
$$
\delta A_\mu = \d_\mu \Lambda,\quad \delta \psi = -\imath e \Lambda \psi, 
\quad \delta \bar{\psi} = \imath e \Lambda \bar{\psi},
$$
where $\Lambda=\Lambda(x)$ is considered to be small. 
Under this variation the integrand in the functional integral 
\eqref{qed-gf}, after integration by parts, acquires a multiplicative factor $e^{\delta_\Lambda}$, with 
\begin{equation}
\delta_\Lambda \equiv \int d^d x \left[ 
-\frac{1}{\alpha} \d^2 (\d_\mu A_\mu) + \d_\mu J_\mu + e(\bar{\psi}\chi - \bar{\chi}\psi) 
\right] \Lambda(x).
\end{equation}
Considering $\delta_\Lambda$ as small we can approximate $e^{\delta_\Lambda} \approx 1 + \delta_\Lambda$ and proceed with the derivation 
procedure from 
\begin{equation}
\delta = \langle \delta_\Lambda \rangle = 0 \label{agp}
\end{equation}

The standard procedure of the variation of action 
with a gauge fixing term \cite{Ryder1985} with respect to $\Lambda_a(x)$ \eqref{agp} leads to the equations \cite{AA2009}: 
\begin{widetext}
\begin{align}
q_\mu \Gamma_{\mu a_4 a_3 a_1}(p,q,p+q) &=& \int \frac{da_2}{a_2}
G_{a_1 a_2}^{-1}(p+q) \tilde M_{a_2a_3a_4}(p+q,q,p) \label{wtes}
- \int \frac{da_2}{a_2} \tilde M_{a_1a_3a_2}(p+q,q,p) G_{a_2 a_4}^{-1}(p), \\
\nonumber \hbox{where\ }  
& &
\tilde M_{a_1 a_2 a_3}(k_1,k_2,k_3) = (2\pi)^d \delta^d(k_1-k_2-k_3) \overline{\tilde g}(a_1k_1)
\tilde g(a_2 k_2) \tilde g(a_3 k_3).
\end{align}
\end{widetext}
The equation \eqref{wtes} is  exactly the wavelet transform  of the standard Ward-Takahashi identity  \eqref{wt1}.
%%%%%%%%%%%%%%%%%%%%%%%%%%%
\subsection{QCD example}
Same as in QED we can evaluate the gluon vacuum polarization 
operator using $g_1$ as the basic wavelet. The corresponding 
one-loop diagram is shown in Diagram \ref{vpg}.
\begin{align}\label{vpg}
\feyn{\vertexlabel^{A,p}g f0 \vertexlabel^{C,p+l}gl  \vertexlabel_{D,l} glu f0 g \vertexlabel^{B,p}} \equiv \Pi^{(\cA)}_{AB,\mu\nu}(p) = \\[1cm]
\nonumber = -\frac{g^2}{2} f^{ABC} f^{BDC} \int \dk{l}{4} \frac{N_{\mu\nu}(l,p)F_{\cA}(l+p,l)}{l^2 (l+p)^2},
\end{align}
where 
\begin{align*}
N_{\mu\nu}(l,p)&=& 10 l_\mu l_\nu + 5(l_\mu p_\nu + l_\nu p_\mu) 
-2 p_\mu p_\nu \\ &+& (p-l)^2 \delta_{\mu\nu} + (2p+l)^2 \delta_{\mu\nu}
\end{align*}
is the tensor structure of the vacuum polarization diagram 
\eqref{vpg} in $\R^4$ Euclidean space. $\cA$ is the minimal 
scale of two external lines. The regularizing function, 
if calculated with $g_1$ wavelet, has the form \eqref{FA}:
$$F_{\cA}(l+p,l) =\exp(-2\cA^2 (l+p)^2 -2\cA^2 l^2).$$

Symmetrizing the loop momenta in equation \eqref{vpg} by 
substitution 
$
l = q - \frac{p}{2},$
we obtain 
\begin{align}\nonumber 
 \Pi^{(\cA)}_{AB,\mu\nu}(p) &=& -\frac{g^2}{2} f^{ACD} f^{BDC} 
 \int \dk{q}{4}F_{\cA}(p,q) \times \\
& & \frac{10 q_\mu q_\nu - \frac{9}{2}p_\mu p_\nu + \delta_{\mu\nu}(\frac{9}{2}p^2+2q^2)
 }{
\left[q^2 - \frac{p^2}{4} \right]^2 
 } \label{pg1g}
 \end{align}
 For $g_1$ wavelet the regularizing function $F_{\cA}(p,q)$ 
 is given by equation \eqref{FA}.
 
 The integral \eqref{pg1g} can be easily evaluated in infrared 
 limit where ordinary QCD is divergent: 
 \begin{widetext}
 \begin{equation}
 \Pi^{(\cA,g_1)}_{AB,\mu\nu}(p\to0) = -g^2 f^{ACD} f^{BDC} 
 \int \dk{q}{4} \frac{e^{-4 \cA^2 q^2}}{q^4} 
[5 q_\mu q_\nu +q^2 \delta_{\mu\nu}].
  \end{equation}
  \end{widetext}
 Making use of the isotropy 
 $$
 d^4 q \to 2\pi^2 q^3 dq, \quad q_\mu q_\nu \to \delta_{\mu\nu}
 \frac{q^2}{4}
 $$
 we get 
 \begin{align*}
 \Pi^{(\cA,g_1)}_{AB,\mu\nu}(p\to0) = 
 -\frac{
 9 g^2 f^{ACD} f^{BDC} \delta_{\mu\nu} 
 }{32} 
 \int_0^\infty q dq e^{-4 \cA^2 q^2}  \\ 
= -\frac{
9 g^2 f^{ACD} f^{BDC}\delta_{\mu\nu}
}{256 \cA^2}. 
 \end{align*}
Similar contribution comes from the ghost loop.
%%%%%%%%%%%%%%%%%%%%%%%%%%%%%%%%%%%%%%%%%%%%%%%%%%%%%%%%%%%%%%
\section{Conclusion}

In this paper we developed a regularization 
method for quantum field theory based on a continuous 
wavelet transform. Regardless significant amount 
of works devoted to wavelet-based regularization 
in different quantum field theory models \cite{BF1987,Federbush1995,Battle1999},
all those are basically the lattice theories. The novelty of 
the present approach, developed by the authors \cite{Alt2002G24J,Altaisky2010PRD,AK2011}, consists in using 
continuous wavelet transform to substitute the local  
fields $\phi(x)$ by the scale-dependent fields $\phi_a(x)$, 
defined as wavelet-coefficients of the physical field. 
Substitution of such fields into the action, supplied by appropriate causality assumptions and operator ordering 
\cite{CC2005,AltaiskyPEPAN2005,Altaisky2010PRD}, results 
in effective regularization of Feynman graphs, which makes 
each internal line decay as an effective 
factor $\propto e^{-p^2 A^2}$, where $A$ is the minimal 
scale of all internal lines, and $p$ is momentum. 

Regulazation factors, that are technically similar to 
our approach, were already known in QCD.
They are related to the modification of the gluon 
vacuum state to the instanton vacuum, with the parameter 
$A$ understood as the size of the instanton \cite{DB2002,Dorokhov2005}. The difference between the 
instanton vacuum model and our model is that the 
scattered quark fields are local fields in the instanton 
model and only the interaction with instanton vacuum is 
smeared. In our approach the incident particles are nonlocal 
wave packets and only the integration over all scales makes the theory local. 

The physics  of using scale-dependent fields $\phi_a(x)$ 
instead of local fields $\phi(x)$ lies in the fact, that 
no physical quantity can be measured in a point, but in 
a region of nonzero size $a>0$. Thus only the finite resolution projections $\phi_a(x)$  of a quantum field $\phi$ are physically meaningful. The $n$-point Green functions for 
such fields constructed by our method are finite by 
construction and do not require regularization. The gauge 
invariance of the theory results in appropriate Ward-Takahashi 
identities, which are the projections of ordinary Ward-Takahashi identities 
onto finite resolution spaces.

The practical applications of our approach can be found 
in such physical settings where the separation of the field 
to the components of different scales is physically meaningful.
Such models have been presented in QED calculations of 
the dependence of the Casimir force on the size of displacement 
in measurement \cite{AK2011}, and also in application of 
quantum field theory methods to the calculation of correlations 
of the turbulent velocity fluctuations of different scales 
\cite{Altaisky2006DAN}. We strongly hope that, regardless the yet unsolved problem of deriving renormalization group equation in continuous limit 
of wavelet-based theory, this method can be also 
applied for QCD calculations, where it was originally 
proposed \cite{Federbush1995}. 
 
\begin{acknowledgments}
The research was supported in part by RFBR Project 13-07-00409 and by the Program of Creation and Development of the National University of Science and Technology "MISiS". The authors are thankful to A.E.Dorokhov, V.G.Kadyshevsky, D.I.Kazakov and O.V.Teryaev for useful discussions, and to anonymous Referee for a number of valuable comments.

\end{acknowledgments}
\appendix
\section{Dirac $\gamma$-matrices in Euclidean space}
\begin{align}
\gamma_\mu \gamma_\nu + \gamma_\nu \gamma_\mu = -2\delta_{\mu\nu}
\\
\gamma_\mu\gamma_\mu =-4, \quad \gamma_\mu \slashed{p} \gamma_\mu = 2 \slashed{p}
\end{align}
Slashed vectors denote convolution with Dirac gamma-matrices  $\slashed{k}=\gamma_\mu k_\mu$, 
$\slashed{k}\slashed{k}=-k^2$
\section{Feynman rules in Euclidean space}
Photon propagator is taken in Feynman gauge:
$$
D(k) = \frac{\delta_{\mu\nu}}{k^2}$$

Fermion propagator 
$$G^{(2)}_E(p) = \frac{-\imath}{\slashed{p}+m} = \imath \frac{\slashed{p}-m}{p^2+m^2}
$$

Electron-fermion vertex: $$-\imath e \gamma_\rho$$

Besides that each fermion vertex results in extra sign $-$ of the whole 
diagram.  

\section{Functions and Integrals \label{gfi:app}}
Exponential integral of the first type 
$$\Ei_1(z)=\int_1^\infty \frac{e^{-xz}}{x}dx$$ 
Integrals for angle integration in Euclidean Green functions 
\cite{AltSIGMA07}:
\begin{align}
\nonumber I_k(y) &\equiv& \int_0^\pi d\theta \frac{\sin^{2}\theta \cos^{2k}\theta}{\beta^2(y)-\cos^2\theta}, \\ 
          I_0(y) &=& \pi (1-\sqrt{1 - \beta^{-2}(y)}),\label{Ik} \\
\nonumber I_1(y) &=& -\frac{\pi}{2} + \beta^2(y)I_0(y), \\
\nonumber        &\ldots &  
\end{align}
The constants $A,B$ for the vacuum polarization 
diagram \eqref{padef} are given by 
$ 4A+B=1,\quad A+B=I_1/I_0$, from where we get 
$$
A = \frac{1}{3} + \frac{\pi}{6} I_0^{-1}(y) -\frac{1}{3}\beta^2(y),  
B = -\frac{1}{3} -\frac{2\pi}{3} I_0^{-1}(y)+\frac{4}{3}\beta^2(y)  
.$$
Integrals in one-loop fermion-photon vertex 
\begin{align}
J^{(0)} = \int \dk{f}{4} \frac{F_A(p-f)F_A(k-f)F_A(f)}{
[(p-f)^2+m^2][(k-f)^2+m^2] f^2
} \\
J^{(1)}_\mu = \int \dk{f}{4} \frac{f_\mu F_A(p-f)F_A(k-f)F_A(f)}{
[(p-f)^2+m^2][(k-f)^2+m^2] f^2
} \\
J^{(2)}_{\mu\nu} = \int \dk{f}{4} \frac{f_\mu f_\nu F_A(p-f)F_A(k-f)F_A(f)}{
[(p-f)^2+m^2][(k-f)^2+m^2] f^2
} 
\end{align}
%%%%%%%%%%%%%%%%%%%%%%%%%%%%%55
% g24.eps, tnew2.eps, mink1.eps, mink2.eps, g1.eps, 
% sa.eps, vpd.eps, rc1.eps 
%\bibliography{qft}
%\include{gau3r2.bbl}

\end{document}